\documentstyle[                  
preprint,                  
aps]{revtex}                  
\begin{document}                  
\draft                  
                  
\title{The size of electron-hole pairs in $\pi$ conjugated systems                   
}                  
                  
\author{M. Knupfer, T. Pichler, M. S. Golden, and J. Fink}                  
\address{                  
Institut f\"ur Festk\"orper- und Werkstofforschung Dresden, D-01171                  
Dresden, Germany                  
}                  
                  
\author{M. Murgia, R. H. Michel, R. Zamboni, and C. Taliani}                  
\address{                  
Istituto di Spettroscopia Moleculare, Consiglio Nazionale delle              
Ricerche, Via P. Gobetti 101, 40129 Bologna, Italy}                  
                  
\maketitle                  
\date{\today}                  
                  
\begin{abstract}                   
We have performed momentum dependent electron       
energy-loss studies of the electronic excitations in              
sexithiophene and compared the results to those from parent oligomers.            
Our experiment probes the dynamic structure factor $S({\bf q},\omega)$ and        
we show that the momentum dependent intensity              
variation of the excitations observed can be used to extract       
the size of the electron-hole pair created in the              
excitation process. The extension of the electron-hole pairs       
along the molecules is comparable to the length of the             
molecules and thus maybe only limited by structural constraints.            
Consequently, the primary     
intramolecular electron-hole pairs are relatively weakly bound.       
We find no evidence for the formation of        
excitations localized on single thiophene units.             
\end{abstract}                  
                  
                  
There is still growing interest in the physical properties       
of various $\pi$ conjugated systems                 
because many polymers and oligomers,       
have become promising candidates for                  
applications in devices such as organic light emitting       
diodes \cite{burroughes90,greenham95}, field effect                 
transistors \cite{greenham95,horowitz96}, optical converters \cite{fichou94}       
or molecular switches \cite{tsivgoulis97}. In                  
addition, $\pi$ conjugated molecules with a finite, well defined chain       
length can serve as model systems for the                  
investigation of general and fundamental properties of the whole material class.      
Moreover, $\pi$-conjugated systems bridge molecular and extended electronic      
states and therefore also allow one to study special cases of Heisenberg,      
Hubbard or other models for narrow band solids.       
       
\par       
Any potential application in electronic or optical devices       
and their optimization requires an understanding of                  
the electronic structure of the system in question. Hereby,       
the excited electronic states are of particular interest                  
in polymers or oligomers as they are directly related to       
processes such as light absorption and emission,        
photoconductivity or exciton                  
formation. One of the remaining questions is to what extent       
exciton formation plays a role in the excited state                
and how large is the related exciton binding energy.       
Exciton binding energies ranging from 0.1 to more than 1                
eV have been discussed previously                
\cite{friend87,lee93,hagler94,leng94,kersting94,blinov95,bredas96,knupfer97}.       
Directly connected with the                
exciton formation and binding energy is the spatial extension of a possible       
excitonic state, i.e. the size of the                
electron-hole pair formed in the excitation process.                   
                
\par       
In this contribution we present a detailed analysis of       
optically allowed and forbidden electronic excitations of             
higly oriented sexithiophene ($\alpha$-6T) thin films       
which have been studied using electron energy-loss                  
spectroscopy (EELS) in transmission. The results are compared       
to those from parent $\pi$ conjugated oligomers.               
EELS is a measure of the dynamic structure factor $S({\bf q},\omega)$ and thus      
probes the form factor of the electronic excitation in question.      
We demonstrate that the momentum-dependent intensity variation of                  
different electronic excitations can be used to extract the size (or separation)       
of the electron-hole pair created in        
the corresponding excitation process which then additionally gives an       
estimate for the electron-hole binding                 
energy.   
To our knowledge, this represents the first determination of the   
spatial extension of electronic excitations via a momentum dependent   
study of $S({\bf q},\omega)$.  
Previously, those data were only analysed on the basis of a quasi band  
structure which completely neglects excitonic effects \cite{pellegrin91,zojer97}.  
Our results argue against the intrinsic formation of strongly       
confined excitons (i.e. excitons that are spatially confined to a single      
thiophene or even smaller unit) in $\pi$             
conjugated systems               
by light absorption or inelastic electron scattering.                  
                
\par       
Thin, crystalline $\alpha$-6T films ($\sim$ 1000 \AA)       
have been grown by evaporation onto a KBr (001) single            
crystal under UHV conditions. The evaporation rate was       
0.1 \AA/s and the substrate was held at room            
temperature. Electron diffraction and direction dependent       
EELS studies showed that the films were highly            
oriented such that the momentum transfer in our measurements could be      
aligned predominantly parallel to the long axis of the $\alpha$-6T molecules.     
The EELS measurements were carried out using a 170 keV         
spectrometer described elsewhere \cite{fink89}.         
We note that at this high primary beam energy only singlet excitations are                   
possible. The energy and momentum resolution were chosen                   
to be 120 meV and 0.05 \AA$^{-1}$, respectively.       
The loss function $Im(-1/\epsilon({\bf q},\omega$)), which is        
proportional to the dynamic structure factor $S({\bf q},\omega)$, has         
been measured for various momentum transfers, $q$.      
Thus, our experiment is       
equivalent to polarization dependent optical absorption            
studies with the light polarization vector being parallel       
to the molecular axis, but it additionally allows one to            
study the excitation properties of the electronic system as       
a function of $q$. The measured loss functions       
have been corrected for contributions from the            
elastic line and multiple                  
scattering \cite{fink89}. The optical conductivity,              
$\sigma({\bf q},\omega) = \omega\epsilon_0Im(\epsilon({\bf q},\omega))$,       
which is a measure for the single particle excitations       
has been derived performing a Kramers-Kronig analysis of       
the loss function. The absolute value of the loss function       
at low momentum transfer was determined using the refractive index $n$            
$\sim$ 2.2 \cite{oelkrug96}, at higher momentum transfers the       
oscillator sum rule was applied \cite{fink89}.                  
                
\par       
In Fig. 1 we present the optical conductivity,       
$\sigma$, of $\alpha$-6T for various momentum transfers parallel                  
to the molecule axis. Strong variations with increasing momentum transfer are visible.                 
At $q$ = 0.15 \AA$^{-1}$ $\sigma$ has a strong       
maximum at about 2.7 eV (feature I) with weak shoulders at              
about 3.2 and 3.7 eV (features II and III) followed       
by a small peak centered at about 4.4 eV (feature IV). The                 
spectrum is fully consistent with those from optical       
absorption studies \cite{oelkrug96,egelhaaf95} or EELS            
measurements in reflexion \cite{egelhaaf95} of crystalline       
$\alpha$-6T films. The fine structure of the first absorption feature             
which has been observed in optical studies and has been       
associated with phonon satellites \cite{muccini97} or            
exciton formation \cite{blinov95} is scarcely resolved in       
our data. Only low energy shoulders at about 2.0 and        
2.4 eV can be seen.      
These have been assigned to aggregates \cite{defect} and the      
lowest Davydov component of the 1$^1$B$_u$ molecular      
excitation \cite{davy}, respectively.     
An energetically higher lying Davydov component appears at     
2.7 eV in optical measurements \cite{davy} also consistent with     
the EELS spectra shown in Fig. 1.    
In the following, we will not consider       
those fine structures but we will take the intensity        
of the features I - IV as a measure for the strength of       
the corresponding electronic excitation as a whole. This is        
justified by the fact that these features do not change       
in line shape or energy position as a function of $q$,            
e.g. the low energy shoulders visible in Fig. 1     
do not show an intensity       
variation different from the main part of feature I (see also        
discussion below). Following the spectra in Fig. 1 with       
increasing momentum transfer, feature I decreases in        
intensity whilst features II - IV increase successively       
before they also start to decrease or saturate.               
                 
\par       
In order to obtain the momentum dependent intensity       
variation of the electronic excitations as                  
observed in Fig. 1, we have modelled the optical       
conductivity with a sum of Lorentz oscillators:                 
                 
\begin{equation}                  
\sigma(\omega) = \epsilon_0\sum_{j}\frac{\omega^2f_j\gamma_j}                  
{(\omega_j^2-\omega^2)^2+\omega^2\gamma_j^2},   \label{G1}                  
\end{equation}                  
                 
with $\omega_j$ being the energy position, $\gamma_j$ the width       
and $f_j$ the oscillator strength of the                  
corresponding excitation. The result of this fit for the       
momentum dependent oscillator strength (or intensity) of                  
the features I - IV is shown in Fig. 2. The momentum position of       
the intensity maximum clearly shifts to higher                  
momentum transfers on going from excitation I to IV. Such a       
behavior has also been observed for $\beta$- carotene \cite{pellegrin91}  
and hexaphenyl \cite{zojer97}, which       
are further representatives of $\pi$               
conjugated oligomers. The momentum position, $q_{max}$,       
of the intensity maxima of the first electronic        
excitations of $\alpha$-6T, hexaphenyl and       
$\beta$-carotene are summarized in Table I.                  
                
\par       
In the remainder of this paper we show that the       
observed intensity variations can be rationalized considering                  
the Taylor expansion of the transition matrix       
element and that one can directly derive a mean extension of the             
electron-hole pairs created in the excitations process.       
The matrix element, $M$, for EELS is proportional to                 
$<f \mid exp(iqr) \mid i>$ which can be expanded to                 
                 
\begin{equation}                  
M \propto \sum_{n}\frac{i^n}{n!} (q<r>)^n <f \mid (\frac{r}{<r>})^n \mid i>.   \label{G2}                  
\end{equation}                 
                 
Hereby, the introduction of a mean radius $<r>$ allows one to    
separate the characteristic dimensionless $(q<r>)$-dependence    
of the matrix element from the (now also dimensionless) $q$-independent excitation    
probability $<f \mid (r$/$<r>)^n \mid i>$. In the case of excitations    
with a {\it specific} multipole character (e.g. dipole excitations) the latter has    
a finite value only for the corresponding $n$ (e.g. $n = 1$).
Thus, the momentum dependence of the excitation intensity $I_n$ ($\propto$ $|M|^2$) of                  
an excitation with a {\it specific} multipole character can be written as                 
                 
\begin{equation}                  
I_n \propto \frac{n!^{-2}(q<r>)^{2n}}{N}, \hspace{0.8cm} N = \sum_{n}\frac{(q<r>)^{2n}}{n!^2}.                     
\label{G3}                  
\end{equation}                 
                  
$N$ is a sum over the intensities of all excited (final) multipole contributions and represents a 
normalization factor which guarantees the oscillator strength sum rule.                  
In Fig. 3 we show the intensities $I_n$ as a       
function of $q<r>$ for $n$ = 1 to 4. Maxima are found at               
$q<r>$ = 0, 2.2, 3.2 and 4 for $n$ = 1, 2, 3 and 4, respectively.                

The mean radius $<r>$ as introduced in Eq. 2 gives a measure for the extension     
of the electron-hole wave function $\Psi_{eh}({\bf r})$ in the             
excited state which represents the probability amplitude to find the electron at       
a certain distance $<r>$ assuming that the hole is fixed (${\bf r}$ denotes the relative 
coordinate ${\bf r}$ = ${\bf r}_e$ - ${\bf r}_h$). 
This can be seen from the dynamic structure factor $S({\bf q},\omega)$ which, in analogy to 
the static atomic structure factor $S({\bf q})$, is the Fourier transform of the electron distribution around 
the hole in the excited state, $\mid \Psi_{eh}({\bf r}) \mid^2$. Assuming an electron-hole pair 
with a hydrogen-like $1s$ wave function having a (Bohr) radius of 1 \AA, $S({\bf q},\omega)$ 
adopts a $q$ dependence which is also shown in Fig. 3 (open circles) 
and which is in reasonable agreement with the $q$-dependent intensity variation of a dipole 
excitation with the same mean radius as expected from Eq. 3.  
From Fig. 3 the error of our approximation regarding the mean radius of the 
excited state wave function can be estimated to be of the order of 20 - 25 \%. We 
emphasize that this uncertainty does not affect the conclusions discussed below. 
Additionally, Fig. 3 shows that higher order excitations occur at       
higher momentum transfers. The similarity                 
between the intensity variations observed for $\alpha$-6T (Fig. 2), hexaphenyl \cite{zojer97}              
or $\beta$-carotene \cite{pellegrin91} and those calculated                  
according to Eq. \ref{G3} (Fig. 3) is striking. It strongly       
indicates that the spatial charge distributions of the            
excited states, which in $\alpha$-6T correspond to the       
excitation energies of 3.2, 3.7 and 4.4 eV, are            
predominantly of higher order multipole character.       
Taking this into account, a comparison of the intensity            
maxima predicted by Eq.             
\ref{G3} (see Fig. 3) and those found in the momentum       
dependence of the optical conductivity can be used to            
derive an estimate of the mean radius $<r>$ of the       
excited state wave functions $\Psi_{eh}({\bf r})$.           
Furthermore, the observed intensity of feature I, which       
is of particular interest as it represents the lowest lying,       
optically allowed singlet excitation, reaches its minimum at $q$ $\sim$ 0.55 \AA$^{-1}$.           
Comparing this to the value for $q<r>$ in Fig. 3 where       
the intensity of the dipole allowed transition becomes           
essentially zero ($q<r>$ $\sim$ 4.5) one obtains an estimate       
for $<r>$ of the dipole allowed excitation I of about 8 \AA.       
An equivalent comparison can also be performed for $\beta$-carotene   
\cite{pellegrin91} and hexaphenyl \cite{zojer97}.   
The results for $<r>$ of the first four electronic excitations            
are presented in Table I. We note that in the       
measurements the momentum vector, {\bf q}, is oriented       
predominantly parallel to the long axis of the molecules and               
that we therefore probe the mean radius $<r>$ of the       
excitated state wave functions along the molecules.           
           
\par       
Interestingly, the mean radius $<r>$ of the excited state       
wave functions of the oligomers compared in Table I                 
indicates a total extension of the electron-hole wave function       
($2<r>$) of about 15 - 20 \AA, i.e. the electron-hole pairs       
are spread over the entire molecules independent of the details of the molecular structures.              
Moreover, from our results there is no             
evidence for the formation of a split-off, strongly localized       
excitonic state whose extension would be significantly             
smaller than the values discussed above as there is no low       
energy feature visible whose intensity variation with             
$q$ is different from the main features. This means that              
electron-hole pairs in the $\pi$ conjugated oligomers       
discussed here are only limited by the finite length of       
the corresponding oligomer. The extension of the electron-hole       
pairs additionally gives an estimate of the electron-hole pair binding energy, $E_B$:                
               
\begin{equation}               
E_B \sim \frac{1}{4\pi\epsilon_0\epsilon_r}\;\frac{e^2}{<r>}\;,                
\end{equation}               
               
which is screened by the static dielectric constant       
$\epsilon_r$ ($\sim$ 4.8 for $\alpha$-6T \cite{oelkrug96}).            
Taking into consideration the size of an               
electron-hole pair as discussed above, this simple consideration leads to a                
binding energy $E_B$ of about 0.3 eV.       
This also indicates that electron-hole pairs in $\pi$ conjugated            
systems do not form strongly bound excitons                
with binding energies of 1 eV or larger but they rather are in a weakly bound state.                 
           
\par       
Recent studies of the intrinsic photoconductivity of PPV-type conjugated polymers \cite{barthprl} arrived at             
similar conclusions. The thermalization distance which is equivalent to $<r>$      
was reported to be 10 - 20 \AA\ with an             
electron-hole binding energy of less than 0.4 eV. Theoretical studies \cite{bredasa,bredasb} on the basis of an      
Intermediate Neglect of Differential Overlap (INDO) approach of the excited state      
wave function $\Psi_{eh}({\bf r})$ in PPV-type oligomers also found an extension of      
the first singlet excited state over the entire molecule, in good agreement to our experimental results.      
Additionally, our results agree well with recent            
calculations of excitons in conjugated polymers within the density matrix renormalization group approach     
based            
upon an extended Hubbard model \cite{bomanprb}. These calculations, which take nearest neighbor Coulomb            
interaction into account, predict the electron-hole separation in the dipole-allowed $B_u$ channel to scale with          
the system size, i.e. they also do not find strongly confined and tightly bound excitons.             
           
\par     
Consequently, we conclude that in $\pi$ conjugated systems strongly confined excitations     
($<r>$ of the order of 2 \AA) are not            
intrinsically formed in electronic excitation processes but that the related electron-hole pairs are spread over            
many monomer units             
with their size probably only limited by the molecule length or by structural imperfections. A possible          
appearance of strongly confined excitons must therefore be due to relaxation processes after the primary      
excitation. For instance, structural relaxation could lead to 'polaron excitons' or impurities could trap electron-     
hole pairs and thus result in further localization and a higher binding energy. This emphasizes the importance     
of      
structural properties of polymer or oligomer films and impurities therein for the performance of devices such as      
organic light emitting diodes.             
             
\par                  
\vspace{1cm}                  
\par                  
{\bf Acknowledgement:}                  
\par                  
We thank S.-L. Drechsler, M. Sing, and E. Zojer for fruitful discussions.    
This work has been partly supported by the    
P.F. MASTA II project DEMO and the EU-TMR Project SELOA.              
\par

\newpage                  
 \parindent0cm                  
                  
\begin{figure}                  
\caption{Optical conductivity, $\sigma$, of $\alpha$-6T for various momentum transfers $q$ parallel to the               
molecules. The curves are offset in y-direction.                 
}                  
\end{figure}                  
                  
\begin{figure}                  
\caption{Intensity variation of the first four excitations observed in Fig. 1 as a function of               
momentum transfer,                  
$q$. Feature I: filled squares, feature II: open circles, feature III: filled diamonds, feature IV: open triangles.          
The lines are intended as a guide to the eye.}          
\end{figure}

\begin{figure}                  
\caption{Intensity of the $n$th multipole ecxitation according to Eq. \ref{G3} ($n$=1: solid line, $n$=2:              
dashed line, $n$=3: dotted line, $n$=4: dashed-dotted line). Additionally shown is the $q$ dependence 
of the dynamic structure factor for a hydrogen-like electron-hole excitation with a Bohr radius of 1 \AA\ 
(dashed line with open circles, see also text).}                  
\end{figure}                  
                  
\begin{table}                  
\caption{Momentum position $q_{max}$ (\AA$^{-1}$) of the intensity maximum of the first electronic                  
excitations in $\alpha$-6T (this study), hexaphenyl \protect\cite{zojer97} and $\beta$-carotene (from               
experiment and theory) \protect\cite{pellegrin91} and the mean radius                  
$<r>$ (\AA) of the excited state wave function along the long axis of the molecules (see text).                  
}                  
\begin{tabular}{c|ccccccc}                  
                  
& $q_{max,II}$  & $q_{max,III}$  &                   
$q_{max,IV}$ & $<r>_I$ & $<r>_{II}$ & $<r>_{III}$ & $<r>_{IV}$ \\                  
\hline                  
$\alpha$-6T & 0.3 & 0.43&0.64& 8&7.3 &7.4 & 6.2 \\                  
hexaphenyl & $\sim$ 0.3& $\sim$ 0.4 & $\sim$ 0.6 & $\le$ 9&7.3 &8 &6.6 \\                  
$\beta$-carotene (exp.) & $\sim$ 0.2&$\sim$ 0.3 & $>$ 0.5& 11& 10&10 &$\le$ 8  \\                  
$\beta$-carotene (theo.) & 0.24&0.4 &0.54 & 12&9.1 &8 &7.4 \\                  
\end{tabular}                  
                  
\label{t1}                  
\end{table}                   
                  
\end{document}